\newif\ifShowKeys
\ShowKeysfalse

\documentclass[11pt,a4paper]{article} 
\pdfoutput=1
\usepackage[no-natbib-sort]{my-jheppub}
\usepackage{amsmath, amssymb}

\usepackage{mathpazo}
\usepackage{bm}
\usepackage{environ}
\usepackage{mathrsfs}
\usepackage{array,arydshln}

\usepackage{graphicx,epsfig}
\usepackage{epic}
\usepackage{youngtab}
\usepackage{float}
\usepackage{color}
\definecolor{maroon}{rgb}{0.8,0.3,0.}

\usepackage{slashed}
\usepackage[nodayofweek]{date time}

\ifShowKeys \usepackage[notcite]{showkeys} \fi

\usepackage{hyperref}

\usepackage{aurical}
\usepackage[T1]{fontenc}

\usepackage{framed}
\definecolor{shadecolor}{RGB}{255, 230, 204}
\newenvironment{claim}{\begin{shaded}\noindent\itshape\ignorespaces}{\end{shaded}}

\allowdisplaybreaks

\newcommand{\be}{\begin{equation}}
\newcommand{\ee}{\end{equation}}

\newcommand{\mc}{\mathcal }

\newcommand{\bv}{\begin{verbatim}}
\newcommand{\ev}{\end{verbatim}}

\newcommand{\la}{\label}

\newcommand{\eps}{\epsilon}
\newcommand{\wt}{\widetilde}
\newcommand{\E}{\text{E}}
\newcommand{\inst}{^{\text{inst}}}
\newcommand{\pert}{^{\text{pert}}}

\title{Exact partition functions for the $\Omega$-deformed $\mc N=2^{*}$ $SU(2)$ gauge theory}
\author[a,b]{Matteo Beccaria} 
\author[a,b]{Guido Macorini} 

\abstract{
We study the low energy effective action  of the 
$\Omega$-deformed $\mc N =2^{*}$ $SU(2) $ gauge theory. It depends on the deformation 
parameters $\eps_{1},\eps_{2}$,  the scalar field expectation value $a$, and the hypermultiplet mass $m$.
We explore the plane $(\frac{m}{\eps_{1}}, \frac{\eps_{2}}{\eps_{1}})$ looking for special features
in the multi-instanton contributions to the prepotential, motivated by what happens in the 
Nekrasov-Shatashvili
limit $\eps_{2}\to 0$. We propose a simple  condition on the structure of poles of the 
$k$-instanton prepotential and show that it is admissible at a finite set of points in the above 
plane. At these special points, the prepotential has poles at fixed positions independent on the 
instanton number. Besides and remarkably, both the instanton partition function and 
the full prepotential, including 
the perturbative contribution, may be given in closed form as functions of the scalar expectation value 
$a$ and the modular parameter $q$ appearing in special combinations of Eisenstein series and Dedekind
$\eta$ function. As a byproduct, the modular anomaly equation can be tested at all orders
at these points. We discuss these special features from the point of view of the AGT correspondence 
and provide explicit toroidal 1-blocks in non-trivial closed form. The full list of solutions
with 1, 2, 3, and 4 poles is determined and described in details.
\vfill }

\affiliation[a]{Dipartimento di Matematica e Fisica Ennio De Giorgi,\\
Universit\`a del Salento, Via Arnesano, 73100 Lecce, 
Italy} 

\affiliation[b]{INFN, Via Arnesano, 73100 Lecce, Italy}

\emailAdd{matteo.beccaria@le.infn.it} 

\begin{document}

%\date{\currenttime}
%\begin{flushleft}\boxed{\small{\tt \today \ \ - \ \  \currenttime }}\end{flushleft}

% \begin{flushright}\small{Imperial-TP-AT-2015-{06}}\end{flushright}				% report number

\maketitle
\flushbottom

\section{Introduction and results}

In this paper we consider the $\Omega$-deformed  $\mc N=2^{*}$ $SU(2)$ gauge theory in four 
dimensions and present novel closed expressions for its low energy effective action 
at special values of the deformation parameters. On general grounds, 
before deformation, the effective action of $\mc N=2$ theories 
is computed by the Seiberg-Witten (SW) curve 
\cite{Seiberg:1994rs,Seiberg:1994aj}. It is the sum of a 1-loop perturbative correction and 
an infinite series of  non-perturbative instantonic contributions that are weighted by the instanton counting 
parameter $q=e^{i\,\pi\,\tau}$ where $\tau$ is the 
complexified gauge coupling constant  at low energy. Due to $\mc N=2$ supersymmetry, 
the full effective action may be expressed in terms of the analytic prepotential $\mc F(a, m)$ depending on 
the vacuum expectation value $a$ of the scalar 
in the adjoint gauge multiplet and on the 
mass $m$ of the adjoint hypermultiplet \cite{D'Hoker:1999ft}. 

Instead of applying the SW machinery, 
one may compute the effective action by topological twisting the theory
and exploiting localization on the many-instanton moduli spaces
 \cite{Nekrasov:2002qd,Nekrasov:2003rj,Nakajima:2003uh}. 
Technically, this is made feasible by introducing the so-called $\Omega$-deformation of the theory,
{\em i.e.} a modification breaking 4d Poincar\' e invariance and 
depending on two parameters $\eps_{1}, \eps_{2}$ such that the initial theory is recovered
when $\eps_{1},\eps_{2}\to 0$. The role of the $\Omega$-deformation 
 is that of a complete regulator for the instanton moduli space integration 
\cite{Flume:2002az,Bruzzo:2002xf,Flume:2004rp,Nekrasov:2004vw,Marino:2004cn,
Billo:2009di,Fucito:2009rs,Billo:2010bd}. In this approach, it is natural to introduce 
a well defined 
partition function $Z\inst(\eps_{1},\eps_{2},a,m)$ and its associated 
non-perturbative $\epsilon$-deformed prepotential by means of 
\be
\la{1.1}
F\inst(\eps_1,\eps_2, a, m)= -\epsilon_1\epsilon_2\,\log Z\inst(\eps_1,\eps_2, a, m).
\ee
%In the non-trivial 
%limit $\eps_{1},\eps_{2}\to 0$, we obtain the original prepotential $\mc F(a,m)$, up to the classical 
%tree-level contribution. 
It is well established that the quantity in (\ref{1.1}) is interesting at finite values of the 
deformation parameters $\eps_{1},\eps_{2}$, {\em i.e.} taking seriously the deformed theory.
This is  because the amplitudes appearing in the 
expansion $F\pert+F\inst=\sum_{n,g=0}^\infty F^{(n,g)}\,(\eps_1+\eps_2)^{2n}\,
(\eps_1\,\eps_2)^g$ are related to the genus $g$ partition function of the $\mc N=2$
topological string  
\cite{Antoniadis:1993ze,Antoniadis:2010iq,Krefl:2010fm,
Huang:2010kf,Antoniadis:2013mna,Antoniadis:2013epe,Florakis:2015ied} and satisfy a 
powerful holomorphic anomaly equation \cite{Bershadsky:1993ta,Bershadsky:1993cx,Klemm:2002pa,Huang:2009md}. 
Actually, understanding the  exact dependence on the deformation parameters is an interesting topic
if one wants to resum the above  expansion in higher genus amplitudes.
Clearly, this  issue  is  closely related to the Alday-Gaiotto-Tachikawa (AGT) correspondence 
\cite{Alday:2009aq} mapping  deformed $\mc N=2$  instanton partition functions  to 
conformal blocks of a suitable CFT with assigned worldsheet genus and 
operator insertions. AGT correspondence may be checked by working order by order in the number of 
instantons \cite{Poghossian:2009mk,Fateev:2009aw,Alba:2010qc}. For the 
$\mc N=2^{*}$ $\eps$-deformed $SU(2)$ gauge theory 
the relevant CFT quantity is the one-point conformal block on the 
torus, a deceptively simple object of great interest  
\cite{Fateev:2009me,Poghossian:2009mk,Hadasz:2009db,Fateev:2009aw,Menotti:2010en,Marshakov:2010fx,KashaniPoor:2012wb,Piatek:2013ifa,Kashani-Poor:2014mua,Alkalaev:2016ptm}.

\medskip
The AGT interpretation emphasizes the importance of modular properties in the deformed gauge theory.
Indeed, it is known that  SW  methods can be extended to the case of non-vanishing deformation 
parameters $\eps_1, \eps_{2}$ \cite{Gaiotto:2009we,Mironov:2009ib}
and modular  properties have been clarified in the 
undeformed case  \cite{Minahan:1997if,Billo:2011pr} as well as in presence of the 
deformation  \cite{Huang:2011qx,Huang:2012kn}. The major outcome of these studies
are explicit resummations of the instanton expansion order by order in the large $a$ regime.
The coefficients of the $1/a$ powers are expressed in  terms of 
quasi-modular functions of the torus {\em nome} $q$. This approach
can be pursued in the gauge theory  \cite{Billo:2013fi,Billo:2013jba,Billo:2014bja,Billo:2015ria,Billo:2015jta,Billo:2016zbf,Ashok:2016oyh}, in CFT language by AGT correspondence
 \cite{KashaniPoor:2012wb,Piatek:2013ifa,Kashani-Poor:2013oza,Kashani-Poor:2014mua}, and 
 also in the framework of the semiclassical WKB analysis
 \cite{Mironov:2009dv,Mironov:2009uv,Mironov:2009dv,He:2010xa,He:2010if,Popolitov:2013ria,He:2014yka}.

\medskip
An important simpler setup where these problems may be addressed is the so-called 
Nekrasov-Shatashvili (NS) limit \cite{Nekrasov:2009rc} where one of the two $\epsilon$ parameters
vanishes. In this case, the deformed  theory has an unbroken two dimensional 
$\mc N = 2$ super-Poincar\'e invariance and 
its  supersymmetric vacua  are related to the  eigenstates of a quantum integrable system.
Under this Bethe/gauge map, the non-zero deformation 
parameter $\eps$ plays the role of $\hbar$ in the quantization of a classically integrable system. 
Saddle point methods allow to derive a deformed SW curve \cite{Poghossian:2010pn,Fucito:2011pn}
that can also be analyzed by matrix model methods \cite{Marshakov:2010fx,Mironov:2009dv,Mironov:2009uv,Bourgine:2012gy,Bourgine:2012bv}. In the specific case of the $\mc N=2^{*}$ theory,
the relevant 
integrable system is the elliptic Calogero-Moser system \cite{Nekrasov:2009rc} and the associated
spectral problem reduces to the study of the celebrated Lam\'e equation.
Besides, if the  hypermultiplet mass $m$ is taken to be proportional to $\eps$ with definite special ratios
$\tfrac{m}{\eps} = n+\tfrac{1}{2}$, where $n\in\mathbb{N}$, the spectral problem is $n$-gap.
Remarkable simplifications
occur in the $k$-instanton prepotential contributions $F\inst_{k}$ \cite{Beccaria:2016wop}
that may be obtained by expanding the eigenvalues of a Lam\'e equation in terms of its Floquet exponent.
As a byproduct of this  approach, it is possible to clarify the meaning of the poles that appear
in the $k$-instanton prepotential at special values $a = \mc O(\eps)$ of the vacuum expectation value $a$.
Indeed, the pole singularities turn out to be an artifact of the instanton expansion. 

\medskip
In this paper, we inquire into similar problems when both the deformation parameters are switched on,
{\em i.e.} by going beyond the Nekrasov-Shatashvili limit. In particular, we 
explore the $(\alpha, \beta)$ plane where $\alpha, \beta$ are real parameters entering the scaling relation 
\be
\la{1.2}
m = \alpha\,\eps_{1}, \qquad \eps_{2}=\beta\,\eps_{1}.
\ee
In other words, we keep the hypermultiplet mass to be proportional to one deformation parameter with 
ratio $\alpha$, but $\eps_{1}, \eps_{2}$ are generic ($\beta$ is just a convenient 
replacement of $\eps_{2}$). By dimensional scaling, the prepotential is a function 
$\wt F_{(\alpha, \beta)}(\nu)$
of the combination $\nu~=~2\,a/\eps_{1}$ at the fixed point $(\alpha, \beta)$. \footnote{
We shall systematically add a tilde to quantities that are considered under (\ref{1.2}) and 
expressed in terms of the variable $\nu$.}
The dependence on $q$ is not
written explicitly. After this stage preparation, the claim of this paper is the following
\begin{claim}
There exists a finite set of $N$-poles points $(\alpha, \beta)$ such that 
the $k$-instanton prepotential is a rational function of $\nu$ with  poles at a fixed set of 
positions $\nu\in\{\nu_{1}, \dots, \nu_{N}\}$ independent on $k$.
\end{claim}
This claim is motivated by our previous analysis in the restricted Nekrasov-Shatashvili limit 
\cite{Beccaria:2016wop} and is far from obvious. Most important, it has 
far reaching consequences. At the special $N$-poles points, we show that the instanton
partition function and the perturbative part of the prepotential take the exact form 
\be
\la{1.3}
\boxed{
\begin{split}
\wt Z\inst_{(\alpha, \beta)}(\nu) &= 
\frac{\nu^{2N}+\sum_{n=1}^{N}\nu^{2\,(N-n)}\mc M_{2n}(q)}{(\nu^{2}-\nu^{2}_{1})
\dots(\nu^{2}-\nu^{2}_{N})}\,[q^{-\frac{1}{12}}\,\eta(\tau)]^{2\,(h_{m}-1)},\\
\wt F\pert_{(\alpha, \beta)}(\nu) &= -\beta\,h_{m}\,\log\frac{\nu}{\Lambda}-\beta\log
\prod_{n=1}^{N}\bigg(1-\frac{\nu_{n}^{2}}{\nu^{2}}\bigg),
\quad h_{m} = \frac{(\beta+1)^{2}-4\,\alpha^{2}}{4\,\beta}, 
\end{split}
}
\ee
where $\mc M_{2n}$ is a polynomial in the Eisenstein series $\E_{2}, \E_{4}, \E_{6}$
with total modular degree $2n$ with coefficients depending on $\alpha, \beta$, 
and $h_{m}\in \mathbb N$. The total prepotential is thus 
remarkably simple and reads
\be
\la{1.4}
\boxed{
\wt F_{(\alpha, \beta)}(\nu) = -\beta\,h_{m}\,\log\frac{\nu}{\Lambda}-\beta\,\log\bigg(
1+\sum_{n=1}^{N}
\frac{\mc M_{2n}(q)}{\nu^{2n}}\bigg).
}
\ee
These explicit expression satisfy the modular anomaly equation 
expressing S-duality discussed in 
\cite{Billo:2013fi,Billo:2013jba,Billo:2014bja,Billo:2015ria,Billo:2015jta,Billo:2016zbf}.
By applying the AGT  dictionary, (\ref{1.3}) and (\ref{1.4}) predict  toroidal blocks 
in closed form
at very specific
values of the central charge $c$ and of the inserted operator conformal dimension $h_{m}$ -- 
the perturbative part providing interesting special instances of the 3-point DOZZ Liouville 
correlation function. These results are derived and tested by giving a complete list of all the 
$N\le 4$ poles points. These turns out to be 4, 7, 12, and 11 at $N=1, 2, 3,4$ respectively.
 
 \medskip
 The plan of the paper is the following. In Sec.~(\ref{sec:first}) we determine the 1-pole points
 by a direct inspection of the instanton prepotential contributions. 
 In Sec.~(\ref{sec:back}) we discuss the special features of the instanton partition function 
 at the 1-pole points. The AGT interpretation is analyzed in Sec.~(\ref{sec:agt})
 where we also provide various explicit CFT tests of the proposed partition functions.
 In Sec.~(\ref{sec:pert}) we discuss the perturbative part of the prepotential at the 1-pole points.
In Sec.~(\ref{sec:23}) the analysis is extended to $N$-poles points and the cases $N=2,3$ are
fully classified. Finally, Sec.~(\ref{sec:block}) presents a list of special toroidal blocks.
Various appendices are devoted to additional comments.

\section{Looking for simplicity beyond the Nekrasov-Shatashvili limit}
\la{sec:first}

As discussed in the Introduction, we are interested in the scaling limit (\ref{1.2}).
%\be
%\la{1.2}
%m = \alpha\,\eps_{1}, \qquad \eps_{2}=\beta\,\eps_{1},
%\ee
%where $\alpha$, $\beta$ are real parameters.
The instanton partition function is $Z\inst=Z\inst(\eps_{1}, \eps_{2}, a, m)$ and it is convenient to 
introduce
\be
\wt Z\inst_{(\alpha, \beta)}(\nu) = 
Z\inst\bigg(\eps_{1}, \beta\,\eps_{1}, \frac{\eps_{1}\,\nu}{2}, \alpha\,\eps_{1}\bigg) = 
Z\inst\bigg(1, \beta, \frac{\nu}{2}, \alpha\bigg), 
\ee
where we used dimensional scaling independence to remove $\eps_{1}$.
Similarly, we define 
\be
F\inst = -\eps_{1}\,\eps_{2}\,\log Z\inst, \qquad 
\wt F\inst_{(\alpha, \beta)}(\nu) = -\beta\,\log \wt Z\inst_{(\alpha, \beta)}(\nu).
\ee
We shall omit the explicit $(\alpha, \beta)$ index when obvious. Besides, the partition 
function is even in $\alpha$ and we shall always consider $\alpha>0$.

\bigskip

According the the claim presented in the Introduction, we now 
look for special points $(\alpha, \beta)$  such that 
 the $k$-instanton Nekrasov function takes the form 
\be
\la{2.3}
\wt F\inst_{k}(\nu) = \frac{P_{k}(\nu)}{(\nu^{2}-\nu_{1}^{2})^{k}},
\ee
with a polynomial $P_{k}(\nu)$ and a single pole $\nu_{1}\ge 0$ in the variable $|\nu|$.
The Ansatz (\ref{2.3}) is a non-trivial requirement. It is motivated by the analysis in \cite{Beccaria:2016wop},
but its admissibility is actually one of the results of our investigation. To explore the constraints that 
(\ref{2.3}) imposes, we begin by looking at the simple one-instanton case. \footnote{
The  functions $\wt F\inst_{k}$ may be computed by the beautiful 
Nekrasov formula \cite{Nekrasov:2003rj}. Alternatively, for a gauge algebra
$\mathfrak g\in \{\text{A}_{r}, \text{B}_{r}, \text{C}_{r}, \text{D}_{r}\}$ one can also  
apply the methods described in  
\cite{Nekrasov:2002qd,Nekrasov:2003rj,Bruzzo:2002xf,Fucito:2004ry,Shadchin:2004yx,Marino:2004cn,Billo:2012st}. 
}
For $k=1$ we have the explicit expression
\be
\wt F\inst_{1}(\nu) = -\frac{(2 \alpha -\beta +1) (2 \alpha +\beta -1) \left(4 \alpha ^2+3 \beta ^2+6 \beta -4 \nu ^2+3\right)}{8
   (\beta -\nu +1) (\beta +\nu +1)},
\ee
and there is a simple pole $\nu_{1}=|\beta+1|$.
At the two-instanton level, $k=2$,  the denominator of $\wt F_{2}(\nu)$ turns out to vanish at 
\be
|\nu| = \beta+1 (\text{order 2}), \ \beta+2, \ 2\,\beta+1.
\ee
Special cases occur when one of the poles coincides with those at $\nu_{1}=|\beta+1|$. This happens for 
\be
\la{2.6}
%\begin{aligned}
%\beta+1 &= -(\beta+1)   &\longrightarrow && \ \ \beta &= -1, \\
%\beta+1 &= -(\beta+2)   &\longrightarrow && \beta &= -\frac{3}{2}, \\
%\beta+1 &= -(2\beta+1) &\longrightarrow  && \beta &= -\frac{2}{3}.
%\end{aligned}
\beta = -1, \ -\frac{3}{2}, \ -\frac{2}{3}.
\ee
These values must be analyzed separately. Looking at higher values of $k$ we
identify the only non-trivial cases consistent with (\ref{2.3}) \footnote{Here, trivial means
a constant $\wt F_{k}(\nu)$.} 
\be
\la{2.7}
(\alpha, \beta) =  \left(\frac{7}{4}, -\frac{3}{2}\right), \ \left(\frac{7}{6}, -\frac{2}{3}\right).
\ee
Finally, if $\beta$ is not in the set (\ref{2.6}), one checks that 
$\wt F_{2}$ takes the form  (\ref{2.3}) if 
\be
\la{2.8}
(\alpha, \beta) = \left(\frac{5}{2}, -2\right), \ \left(\frac{5}{4}, -\frac{1}{2}\right).
\ee
Pushing the calculation up to $12$ instantons, we confirm that the points in (\ref{2.7}) and (\ref{2.8})
agree with the Ansatz (\ref{2.3}). Thus, the {\em 1-pole} 
condition (\ref{2.3}) selects the following distinct 4 special points
\be
\la{2.9}
\text{X}_{1} =  \left(\frac{5}{2}, -2\right), \ \ \
\text{X}_{2} = \left(\frac{7}{4}, -\frac{3}{2}\right), \ \ \
\text{X}_{3} = \left(\frac{7}{6}, -\frac{2}{3}\right), \ \ \
\text{X}_{4} = \left(\frac{5}{4}, -\frac{1}{2}\right).
\ee

\subsection{Back to the instanton partition functions}
\la{sec:back}

We could  analyze further the structure of the prepotential in 
(\ref{2.3}) at the special points $\text{X}_{i}$ in (\ref{2.9})
by looking for regularities in the polynomials $P_{k}(\nu)$. However, it is much more 
convenient to go back to the instanton partition function. To see why, let us consider $\text{X}_{1}$ as a
first illustration. We find indeed the simple expansion
\be
\la{3.1}
\begin{split}
\wt Z\inst_{X_{1}}(\nu) &= 1-\frac{4 \left(\nu ^2-7\right) q^2}{\nu ^2-1}+\frac{2 \left(\nu
   ^2-13\right) q^4}{\nu ^2-1}+\frac{8 \left(\nu ^2-19\right) q^6}{\nu
   ^2-1}-\frac{5 \left(\nu ^2-25\right) q^8}{\nu ^2-1}\\
   &-\frac{4 \left(\nu
   ^2-31\right) q^{10}}{\nu ^2-1}-\frac{10 \left(\nu ^2-37\right)
   q^{12}}{\nu ^2-1}+\frac{8 \left(\nu ^2-43\right) q^{14}}{\nu
   ^2-1}+\frac{9 \left(\nu ^2-49\right) q^{16}}{\nu ^2-1}\\
   &+\frac{14
   \left(\nu ^2-61\right) q^{20}}{\nu ^2-1}+\mc O(q^{22}).
\end{split}
\ee
After some educated trial and error, we recognize that (\ref{3.1}) is the expansion of the 
following expression
\be
\la{3.2}
\wt Z\inst_{\text{X}_{1}}(\nu) = \frac{\nu^{2}-\E_{2}(q)}{\nu^{2}-1}\,q^{-\frac{1}{3}}\,
\eta(\tau)^{4},
\ee
where
\be
\eta(\tau) = q^{\frac{1}{12}}\,\prod_{k=1}^{\infty}(1-q^{2k}),\qquad q=e^{i\,\pi\,\tau},
\ee
and $\E_{2}$ is an Eisenstein series. \footnote{
Our convention is 
\be
\notag \begin{split}
\E_{2}(q) &= 1-24\,\sum_{n=1}^{\infty}\frac{n\,q^{n}}{1-q^{n}}, \ \ \ \
\E_{4}(q) = 1+240\,\sum_{n=1}^{\infty}\frac{n^{3}\,q^{n}}{1-q^{n}}, \ \ \ \
\E_{6}(q) = 1-504\,\sum_{n=1}^{\infty}\frac{n^{5}\,q^{n}}{1-q^{n}}.
\end{split}
\ee
Modular properties of these quantities may be found, for instance,  in \cite{koblitz2012introduction}.}
Similar expressions are found at the other three special points. The detailed formulas are 
\begin{align}
\la{3.4}
\wt Z\inst_{\text{X}_{2}}(\nu) &= \frac{4\,\nu^{2}-\E_{2}(q)}{4\,\nu^{2}-1}\,q^{-\frac{1}{6}}\,
\eta(\tau)^{2}, \notag \\
\wt Z\inst_{\text{X}_{3}}(\nu) &= \frac{9\,\nu^{2}-\E_{2}(q)}{9\,\nu^{2}-1}\,q^{-\frac{1}{6}}\,
\eta(\tau)^{2}, \\
\wt Z\inst_{\text{X}_{4}}(\nu) &= \frac{4\,\nu^{2}-\E_{2}(q)}{4\,\nu^{2}-1}\,q^{-\frac{1}{3}}\,
\eta(\tau)^{4}.\notag 
\end{align}
The associated all-instanton Nekrasov functions are 
\begin{align}
\la{3.5}
\wt F\inst_{\text{X}_{1}}(\nu) &= 8\,\log[q^{-\frac{1}{12}}\,\eta(\tau)]+2\,\log\left(
\frac{\nu^{2}-\E_{2}}{\nu^{2}-1}
\right), \notag \\
\wt F\inst_{\text{X}_{2}}(\nu) &= 3\,\log[q^{-\frac{1}{12}}\,\eta(\tau)]+\frac{3}{2}\,\log\left(
\frac{4\,\nu^{2}-\E_{2}}{4\,\nu^{2}-1}
\right), \\
\wt F\inst_{\text{X}_{3}}(\nu) &= \frac{4}{3}\,\log[q^{-\frac{1}{12}}\,\eta(\tau)]+\frac{2}{3}\,\log\left(
\frac{9\,\nu^{2}-\E_{2}}{9\,\nu^{2}-1}
\right), \notag \\
\wt F\inst_{\text{X}_{4}}(\nu) &= 2\,\log[q^{-\frac{1}{12}}\,\eta(\tau)]+\frac{1}{2}\,\log\left(
\frac{4\,\nu^{2}-\E_{2}}{4\,\nu^{2}-1}
\right). \notag
\end{align}
Equations (\ref{3.4}) and (\ref{3.5}) are already remarkable because they are non-trivial closed expressions
for the instanton partition function, or prepotential, at all instanton numbers.
It is clear that it would be nice to provide some clarifying interpretation for this 
features at the special points $\text{X}_{i}$.
In the next section, we shall examine the clues coming from AGT correspondence.

\section{AGT  interpretation}
\la{sec:agt}

According to the AGT correspondence, the instanton partition function of $\mc N=2^{*}$ 
$SU(2)$ gauge theory is  \cite{Alday:2009aq,Fateev:2009aw,Poghossian:2009mk} 
\be
\la{4.1}
Z\inst(q,a,m) =\left[\prod_{k=1}^{\infty}(1-q^{2k})\right]^{-1+2\,h_{m}}\,\mc F^{h}_{h_{m}}(q),
\ee
where $\mc F_{\alpha}^{m}(q)$
is the 1-point toroidal block of the Virasoro algebra of central charge $c = 1+6\,Q^{2}$ 
on a torus whose modulus is $q$, with one operator of dimension $h_{m}$ 
inserted and a primary of dimension $h$ in the intermediate channel.
The precise dictionary in terms of the deformation parameters is 
\be
\la{4.2}
\begin{aligned}
b &= \sqrt{\eps_{2}/\eps_{1}},  & \qquad Q &= b+b^{-1}, \\
h_{m} &= \frac{Q^{2}}{4}-\frac{m^{2}}{\eps_{1}\,\eps_{2}}, & \qquad
h &= \frac{Q^{2}}{4}-\frac{a^{2}}{\eps_{1}\,\eps_{2}}.
\end{aligned}
\ee

Assuming the scaling relations (\ref{1.2}), the expressions in  (\ref{4.2}) read
\be
\la{4.3}
\begin{aligned}
b &= \sqrt\beta,  & \qquad Q &= \beta^{\frac{1}{2}}+\beta^{-\frac{1}{2}}, \\
h_{m} &= \frac{(\beta+1)^{2}-4\,\alpha^{2}}{4\,\beta}, & \qquad
h &= \frac{(\beta+1)^{2}\,\eps_{1}^{2}-4\,a^{2}}{4\,\beta\,\eps_{1}^{2}},
\end{aligned}
\ee
with central charge
\be
c = 13+6\,\left(\beta+\frac{1}{\beta}\right).
\ee
In particular, at the four points $\text{X}_{i}$ we obtain the following values for $(c,h_{m})$
\be
\begin{array}{|c|c|c|c|c|}
\hline
& X_{1} & X_{2} & X_{3} & X_{4} \\
\hline
c & -2 & 0 & 0 & -2 \\
h_{m} & 3 & 2 & 2 & 3\\
\hline 
\end{array}
\ee
Of course, points appear in pairs with the same central charge and $\beta$ values related by
$\beta\to \beta^{-1}$. More remarkably, the associated values of the parameter $\alpha$ is always such that
$h_{m}$ is a positive integer.
The toroidal block has a universal prefactor $q^{\frac{1}{12}}/\eta(\tau)$
that is its value at $h\to \infty$. Comparing (\ref{4.1}) with (\ref{3.4})
we can write the general form for all four $\text{X}_{i}$ points as
\be
\la{4.6}
\begin{split}
%Z(\alpha, \beta) &= 
%\frac{4\,a^{2}-(1+\beta)^{2}\,\eps_{1}^{2}\,\E_{2}(q)}{4\,a^{2}-(1+\beta)^{2}\,
%\eps_{1}^{2}}\,[q^{-\frac{1}{12}}\,\eta(\tau)]^{\frac{(-2 \alpha +\beta -1) (2 \alpha +\beta -1)}{2 \beta }} ,\\
\wt Z\inst_{(\alpha, \beta)}(\nu) &= 
\frac{\nu^{2}-\nu_{1}^{2}\,\E_{2}(q)}{\nu^{2}-\nu_{1}^{2}}
\,[q^{-\frac{1}{12}}\,\eta(\tau)]^{2\,(h_{m}-1)} ,
\qquad \nu_{1} = |\beta+1|,\\
%\wt F(\nu; \alpha, \beta) &= 
%\frac{4\,\alpha^{2}-(\beta-1)^{2}}{2}\, \log[q^{-\frac{1}{12}}\,\eta(\tau)]-\beta\,\log
%\frac{\nu^{2}-(1+\beta)^{2}\,\E_{2}}{\nu^{2}-(1+\beta)^{2}}, \\
\wt F\inst_{(\alpha, \beta)}(\nu) &= 
-2\,\beta\,(h_{m}-1)\, \log[q^{-\frac{1}{12}}\,\eta(\tau)]-\beta\,\log
\frac{\nu^{2}-\nu_{1}^{2}\,\E_{2}}{\nu^{2}-\nu_{1}^{2}}.
\end{split}
\ee
Correspondingly, 
the net prediction for the toroidal  block at the above central charge and insertion dimension 
is 
%\be
%\la{4.7}
%\begin{split}
%& \mc F^{h}_{h_{m}}(q, c) = \frac{q^{\frac{1}{12}}}{\eta(\tau)}
%%
%%\frac{4\,a^{2}-(1+\beta)^{2}\,\eps_{1}^{2}\,\E_{2}(q)}{4\,a^{2}-(1+\beta)^{2}\,
%%\eps_{1}^{2}}
%\bigg[1+\frac{(\beta+1)^{2}}{4\,\beta\,h}\,(\E_{2}(q)-1)\bigg], \qquad c = 13+6\,\left(\beta+\frac{1}{\beta}\right).
%\end{split}
%\ee
\be
\la{4.7}
\begin{split}
& \mc F^{h}_{h_{m}}(q, c) = \frac{q^{\frac{1}{12}}}{\eta(\tau)}
\bigg[1+\frac{c-1}{24\,h}\,(\E_{2}(q)-1)\bigg], \qquad (c, h_{m}) = (0,2)\ \text{or}\ (-2,3).
\end{split}
\ee
We remark that the above $(c, h_{m})$ may well be pathological for a physical CFT. Nevertheless,
the toroidal block is defined by the Virasoro algebra for abritrary values of $c, h_{m}$, and $h$.
Eq. (\ref{4.7}) must be taken in this sense.
We checked (\ref{4.7}) against Zamolodchikov recursive determination of the toroidal block
\cite{Zamolodchikov:1985ie,Zamolodchikov:1995aa,Perlmutter:2015iya} with perfect agreement. Of course, by AGT, this is same as 
Nekrasov calculation. The remarkably simple form (\ref{4.7}) is clearly consistent with 
general results for the torus block.
For instance, at leading and next-to-leading order and generic operator dimensions we
have \cite{Piatek:2013ifa}
\be
\mc F^{h}_{h_{m}}(c,q) = 1+\mc F_{1}(h, h_{m},c)\,q^{2}+\mc F_{2}(h,h_{m},c)\,q^{4}+\dots,
\ee
where
\begin{align}
\mc F_{1}(h,h_{m},c) &= 1+\frac{h_{m}\,(h_{m}-1)}{2\,h},\notag \\
\mc F_{2}(h,h_{m},c) &= [4\,h\,(2\,c\,h+c+16\,h^{2}-10\,h)]^{-1} \\
&\bigg[
(8\,c\,h+3\,c+128\,h^{2}+56\,h)\,h_{m}^{2}+(-8\,c\,h-2\,c-128\,h^{2})\,h_{m}\notag \\
&+(c+8\,h)\,h_{m}^{4}+(-2\,c-64\,h)\,h_{m}^{3}+16\,c\,h^{2}+8\,c\,h+128\,h^{3}-80\,h^{2}
\bigg].\notag
\end{align}
Thus, 
\be
\begin{split}
\mc F^{h}_{h_{m}=2}(c=0,q) &= 1+\left(1+\frac{1}{h}\right)\,q^{2}
+\left(2+\frac{4}{h}\right)\,q^{4}+\dots, \\
\mc F^{h}_{h_{m}=3}(c=-2,q) &= 1+\left(1+\frac{3}{h}\right)\,q^{2}
+\left(2+\frac{12}{h}\right)\,q^{4}+\dots,
\end{split}
\ee
in full agreement with (\ref{4.7}) for $c=0, -2$. Notice also that (\ref{4.7}) may be written
\be
\la{4.11}
 \frac{q^{\frac{1}{12}}}{\eta(\tau)}
\bigg[1+\frac{c-1}{24\,h}\,(\E_{2}(q)-1)\bigg] = \left(1+\frac{1-c}{2\,h}\,q\,\partial q\right)\,
\frac{q^{\frac{1}{12}}}{\eta(\tau)} = 
\sum_{k=0}^{\infty} \left(1+\frac{1-c}{h}\,k\right)\,P_{k}\,q^{2k},
\ee
where $P_{k}$ are the coefficients of the expansion of $\prod_{k=1}^{\infty}(1-q^{2k})^{-1}$, {\em i.e.}
 the number of unrestricted partitions of $k$ ($n^{m}$ means $m$ copies of $n$)
\be
\begin{split}
P_{1} &= \#\{(1)\}=1, \ P_{2} = \#\{(2), (1^{2})\}=2, \ P_{3}=\#\{(3),(2,1),(1^{3})\}=3, \\
P_{4} &= \#\{(4), (3,1), (2^{2}), (2,1^{2}), (1^{4})\}=5, \\
P_{5} &= \#\{(5), (4,1), (3,2), (3, 1^{2}), (2^{2}, 1), (2, 1^{3}), (1^{5})\} = 7, \ \text{and so on.}
\end{split}
\ee

\subsection{Explicit CFT computations}
\subsubsection*{The $(c,h_{m})=(0,2)$ conformal block}

It is instructive to derive the result (\ref{4.7}) at $c=0$ from a direct CFT calculation. 
\footnote{
CFT at $c=0$ is obviously quite special since the 2-point function of the stress energy
tensor is then zero, so strictly speaking, the theory is not conformal any more (since
the stress-tensor vanishes identically). However, as we remarked, we are considering the toroidal block as 
a well-defined function of $(c, h_{m}, h)$ that may be regarded as the solution to the Zamolodchikov recursion relations. It would be interesting to revisit our calculation in the language of 
logarithmic CFT, see for instance \cite{Flohr:2013dma,Hogervorst:2016itc}.}
In other 
words, we want to show that 
\be
\la{4.13}
\begin{split}
& \mc F^{h}_{h_{m}=2}(q, c=0) = \frac{q^{\frac{1}{12}}}{\eta(\tau)}
\bigg[1-\frac{1}{24\,h}\,(\E_{2}(q)-1)\bigg] = 
\sum_{k=0}^{\infty} \left(1+\frac{1}{h}\,k\right)\,P_{k}\,q^{2k},
\end{split}
\ee
where we used (\ref{4.11}). The toroidal block is obtained as 
\be
\mc F^{h}_{h_{m}}(q, c) = q^{-h+\frac{c}{12}}\,\text{Tr}_{h}\bigg(
q^{L_{0}-\frac{c}{12}}\,\varphi_{h_{m}}(1)\bigg),
\ee
where the trace is over the descendants of $\varphi_{h}$.
The starting point is thus conformal descendant decomposition of the {\em diagonal} part of the 
OPE
\be
\la{4.15}
\begin{split}
\varphi_{h_{m}}(x)\,\varphi_{h}(0) &= \sum_{Y} x^{-h_{m}+|Y|}\,\beta^{Y}\,L_{-Y}\,\varphi_{h}(0)\\
&= x^{-h_{m}}\,(1+x\,\beta^{(1)}\,L_{-1}+x^{2}\,(
\beta^{(2)}\,L_{-2}+\beta^{(1,1)}\,L_{-1}^{2}
)+\dots)\,\varphi_{h}(0),
\end{split}
\ee
where $Y$ denotes a unrestricted partition of $|Y|$ and $L_{n}$ are Virasoro generators
\be
\begin{split}
& Y =\{k_{1}\ge k_{2}\ge \cdots >0\}, \qquad |Y|=k_{1}+k_{2}+\dots., \\
& L_{-Y} = L_{-k_{1}}\,L_{-k_{2}}\,\dots.
\end{split}
\ee
The coefficients $\beta$ in (\ref{4.15}) are determined by conformal symmetry and are functions
of $h, h_{m}, c$. As a consequence of unbroken conformal symmetry ($c=0$) and of the 
fact that $h_{m}=2$
is the same dimension as that of the energy momentum tensor, one finds that the only vanishing 
$\beta$ coefficients are those associated with simple $L_{-n}$ descendants. Besides they are all equal
\be
\beta^{(n)} = \frac{1}{h}, \qquad \beta^{(n,n')} = 0, \qquad \beta^{(n, n', n'')} = 0, \qquad \dots.
\ee
Just to give an example, the explicit $\beta$ coefficients at level 2 are 
\be
\begin{split}
\beta^{(1)} &= \frac{h_{m}}{2\,h}, \qquad 
\beta^{(2)} = \frac{(1+8\,h-3\,h_{m})\,h_{m}}{c\,(1+2\,h)+2\,h\,(8\,h-5)}, \\
\beta^{(1,1)} &= \frac{h_{m}\,(c-16\,h+(c+8\,h)\,h_{m})}{4\,h\,(c\,(1+2\,h)+2\,h\,(8\,h-5))}.
\end{split}
\ee
Computing them at $(c,h_{m})=(0,2)$ we see that indeed
\be
\beta^{(1)} = \beta^{(2)} = \frac{1}{h}, \qquad \beta^{(1,1)}=0.
\ee
Hence, if we apply (\ref{4.15}) to the vacuum, we get  \footnote{Notice that $h_{m}=2$ is not
enough to achieve such simplification. A vanishing central charge is also needed to remove
 descendants with multiple applications of Virasoro operators.}
\be
\la{4.20}
\varphi_{2}(x)\,|h\rangle = \frac{1}{h}\,\sum_{n=0}^{\infty}x^{n-2}\,L_{-n}\,|h\rangle.
\ee
Now, to get the torus block, we need to evaluate the diagonal matrix elements of $\varphi_{2}(x)$.
Using
\be
[L_{n}, \varphi_{2}(x)] = x^{n}\,(x\,\partial+2\,(n+1))\,\varphi_{2}(x),
\ee
we obtain with one index
\be
\begin{split}
\varphi_{2}(x)\,L_{-k}\,|h\rangle &= -[L_{-k}, \varphi_{2}(x)]\,|h\rangle
+L_{-k}\,\varphi_{2}(x)\,|h\rangle\\
&= \frac{1}{h}\sum_{n=0}^{\infty}(2k-n)\,x^{n-k-2}\,L_{-n}|h\rangle
+ \frac{1}{h}\,\sum_{n=0}^{\infty}x^{n-2}\,L_{-k}L_{-n}\,|h\rangle \\
&= \dots+\frac{1}{x^{2}}\,\left(1+\frac{k}{h}\right)\,L_{-k}|h\rangle + \dots,
\end{split}
\ee
where we have shown only the diagonal entry.
Adding one index each time, a similar calculation shows that for any number of indices
\be
\la{4.23}
\begin{split}
\varphi_{2}(x)\,L_{-Y}\,|h\rangle &=  
\dots+\frac{1}{x^{2}}\,\left(1+\frac{|Y|}{h}\right)\,L_{-Y}|h\rangle + \dots.
\end{split}
\ee
Thus the diagonal matrix element of $\varphi_{2}(x)$ associated with the $Y$ descendent 
depends only on $|Y|$. The number of $Y$ with fixed $|Y|$ is the number $P_{|Y|}$
of unrestricted partitions of $|Y|$. Summing over $Y$ with $|Y|=k$ we prove (\ref{4.13}).

\subsubsection*{The $(c,h_{m})=(-2,3)$ conformal block}

A similar computation for $c=-2$ and $h_{m}=3$ is apparently quite less trivial. The main reason is
that the $\beta$ coefficients in the conformal decomposition (\ref{4.15}) do not trivialize in this case.
This complication forbids us to prove the wanted result in general. Nevertheless, we provide an
explicit  check at level 4. Of course, one could simply use the recursive definition of the toroidal block,
but our brute force calculation is perhaps more transparent. Besides, it emphasizes the difference 
compared with the previous $(c,h_{m})=(0,2)$ case.
The starting point is again the OPE (\ref{4.15}) that now takes the following form up to level 4 descendants
\begin{align}
\varphi_{3}(x)&\,\varphi_{h}(0) = 
x^{-3}\,(1+x\,\beta^{(1)}\,L_{-1}+x^{2}\,(
\beta^{(2)}\,L_{-2}+\beta^{(1,1)}\,L_{-1}^{2}
)\\
&+x^{3}\,(
\beta^{(3)}\,L_{-3}+\beta^{(2,1)}\,L_{-2}\,L_{-1}+\beta^{(1,1,1)}\,L_{-1}^{3}
) \notag \\
&+x^{4}\,(
\beta^{(4)}\,L_{-4}+
\beta^{(3,1)}\,L_{-3}\,L_{-1}+
\beta^{(2,2)}\,L_{-2}^{2}+
\beta^{(2,1,1)}\,L_{-2}\,L_{-1}^{2}+
\beta^{(1,1,1,1)}\,L_{-1}^{4}
)
+\dots)\,\varphi_{h}(0) \notag,
\end{align}
with the  simple but non trivial $\beta$ coefficients
\begin{align}
\beta^{(1)} &= \frac{3}{2 h}, \quad
\beta^{(2)} = \frac{12}{8 h+1},\quad
\beta^{(1,1)} = \frac{3}{h (8 h+1)},\quad
\beta^{(3)} = \frac{24 h-1}{2 h (8 h+1)},\notag \\
\beta^{(2,1)} &= \frac{6 h+1}{h^2 (8 h+1)},\quad
\beta^{(1,1,1)} = -\frac{1}{2 h^2 (8 h+1)},\quad
\beta^{(4)} = \frac{3 \left(32 h^2-20 h+1\right)}{h (8 h-3) (8 h+1)},\notag \\
\beta^{(3,1)} &=\frac{3 \left(16 h^2-2 h-1\right)}{h^2 (8 h-3) (8 h+1)},\quad
\beta^{(2,2)} = \frac{24}{(8 h-3) (8 h+1)},\notag \\
\beta^{(2,1,1)} &= -\frac{3 (4 h+1)}{h^2 (8 h-3) (8 h+1)},\quad
\beta^{(1,1,1,1)} = \frac{3}{2 h^2 (8 h-3) (8 h+1)}\notag 
\end{align}
As in (\ref{4.23}), we write
\be
\varphi_{3}(x)\,L_{-Y}|h\rangle = \dots+\frac{1}{x^{3}}\,M_{Y}\,L_{-Y}\,|h\rangle+\dots,
\ee
for certain coefficients $M_{Y}$ functions of $h$. At level 1, we have only 
\be
M_{1} = 1+\frac{3}{h}.
\ee
At level 2, 
\be
\begin{split}
& M_{2} = \frac{8 h+49}{8 h+1}, \quad M_{1,1}=\frac{8 h^2+49 h+12}{h (8 h+1)}, \\
&\sum_{|Y|=2} M_{Y} = 2+\frac{12}{h}.
\end{split}
\ee
At level 3, 
\be
\begin{split}
& M_{3} =\frac{8 h^2+73 h-3}{h (8 h+1)} , \quad 
M_{2,1}=\frac{8 h^2+73 h+3}{h (8 h+1)}, \quad 
M_{1,1,1}=\frac{8 h^2+73 h+27}{h (8 h+1)},\\
&\sum_{|Y|=3} M_{Y} = 3+\frac{27}{h}.
\end{split}
\ee
At level 4 
\be
\begin{split}
& M_{4} = \frac{64 h^3+752 h^2-483 h+24}{h (8 h-3) (8 h+1)}, \quad 
M_{3,1}=\frac{64 h^3+752 h^2-291 h-24}{h (8 h-3) (8 h+1)}, \\
& M_{2,2}=\frac{64 h^2+752 h-99}{(8 h-3) (8 h+1)}, \quad 
M_{2,1,1}=\frac{h+12}{h}, \quad 
M_{1,1,1,1} = \frac{8 h^2+97 h+48}{h (8 h+1)},\\
& \sum_{|Y|=4} M_{Y} = 5+\frac{60}{h}.
\end{split}
\ee
Putting all together we agree with (\ref{4.11}) at $c=-2$. It would be nice to prove the agreement 
at all levels, possibly working in a definite $c=-2$ CFT like the triplet model 
considered in \cite{Flohr:2005cm,Gaberdiel:2006pp}.

\section{Perturbative part of the prepotential at the points $\text{X}_{i}$}
\la{sec:pert}

The prepotential has also a perturbative part, related by AGT to the DOZZ 3-point function in the 
Liouville theory \cite{Dorn:1994xn,Zamolodchikov:1995aa,Teschner:2003en}. The general 
expression for the perturbative part of the  prepotential is ($\wt m = m+\frac{\eps_{1}+\eps_{2}}{2}$)
\cite{Nekrasov:2002qd,Nekrasov:2003rj}
\be
\la{5.1}
\begin{split}
F\pert = \eps_{1}\,\eps_{2}\,\bigg[\gamma_{\eps_{1}, \eps_{2}}(2\,a)+
\gamma_{\eps_{1}, \eps_{2}}(-2\,a)
-\gamma_{\eps_{1}, \eps_{2}}(2\,a+\wt m)
-\gamma_{\eps_{1}, \eps_{2}}(-2\,a+\wt m)
\bigg],
\end{split}
\ee
where 
\be
\gamma_{\eps_{1},\eps_{2}}(x) = \left. \frac{d}{ds}\left(
\frac{\Lambda^{s}}{\Gamma(s)}\,\int_{0}^{\infty}\frac{dt}{t}\,\frac{t^{s}\,e^{-t\,x}}
{(e^{-\eps_{1}\,t}-1)\,(e^{-\eps_{2}\,t}-1)}\right)\right|_{s=0},
\ee
and $\Lambda$ is a renormalization scale. Evaluating $F_{\rm pert}$ by expanding 
at small $\eps_{1}$ and resumming,  we find  that at all $\text{X}_{i}$ points it is possible to write
\be
F\pert = \frac{4\,\alpha^{2}-(\beta+1)^{2}}{4}\,\eps_{1}^{2}\,\log\frac{2\,a}{\Lambda}
-\beta\,\eps_{1}^{2}\,\log\left[1-(1+\beta)^{2}\,\frac{\eps_{1}^{2}}{4\,a^{2}}\right].
\ee
Again, this appears to be a special feature of the $\text{X}_{i}$ points because it is not possible to 
give such a simple expression for $F\pert$ at generic $\eps_{1}, \eps_{2}$ from (\ref{5.1}).
With a redefinition of the UV cutoff, this may be written in the following suggestive form
that we shall generalize later
\be
\la{5.4}
\wt F\pert(\nu) = -\beta\,h_{m}\,\log\frac{\nu}{\Lambda}
-\beta\,\log\left(1-\frac{\nu_{1}^{2}}{\nu^{2}}\right).
\ee

%\subsection{Check of the modular anomaly equation \red{dopo ???}}
%
%Adding the perturbative part of the partition function (see 2.19 of \cite{Billo:2013fi})
%and dropping terms independent on $a$, 
%we obtain the total 
%\be
%\la{6.5}
%\wt F_{\rm tot} = 
%\wt F_{\rm pert}+\wt F = h_{0}\log\frac{\nu}{\Lambda}-\beta\log\left(
%1-(1+\beta)^{2}\,\frac{\E_{2}}{\nu^{2}}
%\right).
%\ee
%where
%\be
%h_{0} = -\beta\,h_{m}=\frac{4 \alpha^{2} -(\beta +1)^{2}}{4}.
%\ee
%Expanding at large $\nu$
%\be
%\wt F_{\rm tot} 
%= h_{0}\log\frac{\nu}{\Lambda}-
%\sum_{\ell=1}^{\infty}\frac{h_{\ell}}{2^{1-\ell}\,\ell}\frac{1}{\nu^{2\ell}},
%\ee
%with 
%\be
%h_{\ell\ge 1} =-2^{1-\ell}\,\beta\,(\beta+1)^{2\,\ell}\,\E_{2}^{\ell}.
%\ee
%One can check that the modular anomaly equation  (see 2.36 of \cite{Billo:2013fi})
%\be
%\frac{\partial h_{\ell}}{\partial \E_{2}} = \frac{\ell}{12}\,\sum_{n=0}^{\ell-1}h_{n}\,h_{\ell-1-n}
%+\beta\,\frac{\ell\,(2\,\ell-1)}{12}\,h_{\ell-1},
%\ee
%is satisfied if 
%\be
%(\beta+2)(2\,\beta+1)(2\,\beta+3)(3\,\beta+2)=0, \qquad\text{and}\qquad |\alpha| = \frac{1}{2}\,
%(13+20\,\beta+13\,\beta^{2})^{1/2}.
%\ee
%and, consistently,  this gives again the four points $\text{X}_{i}$.

\section{Full prepotential and generalization to $N$-poles points}
\la{sec:23}

If we combine the perturbative (\ref{5.4}) and instanton (\ref{4.6}) parts of the prepotential, we 
obtain the remarkably simple expression
\be
\la{6.1}
\wt F = \wt F\pert+\wt F\inst = -\beta\,h_{m}\,\log\frac{\nu}{\Lambda}-\beta\,\log\bigg(
1+\frac{\gamma\,\E_{2}(q)}{\nu^{2}}\bigg),
\ee
with a certain coefficient $\gamma$. This suggests  that it is convenient to organize the total 
prepotential in the form 
\be
\la{6.2}
\wt F = -\beta\,h_{m}\,\log\frac{\nu}{\Lambda}-\beta\,\log\bigg(
1+\sum_{n=1}^{\infty}
\frac{\mc M_{2n}(q)}{\nu^{2n}}\bigg),
\ee
where $M_{2n}(q)$ is a polynomial in $\E_{2,4,6}$ of (quasi-) modular degree $2n$. We emphasize again 
that the 
Ansatz (\ref{6.2}) is  non trivial because $\wt F_{\rm tot}$ is a combination of 
the perturbative and instanton contributions. Our claim is that (\ref{6.2}) can be truncated at maximum
degree $n=N$ for a special set of points  $(\alpha, \beta)$. At such points, the instanton 
partition function takes the special form, see (\ref{1.3})
\be
\la{6.3}
\wt Z\inst_{(\alpha, \beta)}(\nu) = 
\frac{\nu^{2N}+\sum_{n=1}^{N}\nu^{2\,(N-n)}\mc M_{2n}(q)}{(\nu^{2}-\nu^{2}_{1})
\dots(\nu^{2}-\nu^{2}_{N})}\,[q^{-\frac{1}{12}}\,\eta(\tau)]^{2\,(h_{m}-1)} .
\ee
To see how this works in practice, let us parametrize
\be
\begin{split}
\mc M_{2} &= \gamma_{2}\,\E_{2}, \qquad 
\mc M_{4} = \gamma_{4}\,\E_{4}+\gamma_{2,2}\,\E_{2}^{2}, \\
\mc M_{6} &= \gamma_{6}\,\E_{6}+\gamma_{2,4}\,\E_{2}\,\E_{4}+\gamma_{2,2,2}\,\E_{2}^{3}, \\
\mc M_{8} &= \gamma_{2,6}\,\E_{2}\,\E_{6}+\gamma_{4,4}\,\E_{4}^{2}+\gamma_{2,2,4}\,
\E_{2}^{2}\,\E_{4}+\gamma_{2,2,2,2}\,\E_{2}^{4}.
\end{split}
\ee
Replacing these combinations in (\ref{6.3}) and comparing with the Nekrasov function at 5 instanton we obtain explicit 
expressions for the $\gamma$ coefficients  in terms of $\alpha, \beta$. 
%\red{Notice that we do not need the explicit positions
%of the poles $\nu_{1, \dots, N}$. \footnote{\red{
%The poles $\nu_{1},\dots, \nu_{N}$ are naively at known positions $|\beta+1|, |\beta+2|, \dots$, 
%but there can be 
%accidental modifications like for $\alpha=\frac{11}{4}$ and $\beta=-\frac{3}{2}$. Inspection of the first 
%Nekrasov functions gives immediately their positions.}}
%Indeed, it is enough to parametrize the large $\nu$ expansion of the 
%perturbative part as $C_{1}/\nu^{2}+C_{2}/\nu^{4}+\dots$ and determine $C_{i}$ together with the 
%$\gamma$'s.}
The first coefficients are 
\be
\begin{split}
\gamma_{4} &= -\frac{1}{23040 \beta }\,
(-2 \alpha +\beta -1) (-2 \alpha +\beta +1) (2 \alpha +\beta -1) (2 \alpha
   +\beta +1) \\
   &\left(-4 \alpha ^2+13 \beta ^2+20 \beta +13\right), \\
   \gamma_{2,2} &= \frac{1}{73728 \beta ^2}\,(-2 \alpha +\beta -1) (-2 \alpha +\beta +1) (2 \alpha +\beta -1) (2 \alpha+\beta +1)\\
   & \qquad\qquad \left(-4 \alpha ^2+\beta ^2-10 \beta +1\right) \left(-4 \alpha ^2+\beta^2-6 \beta +1\right).
\end{split}
\ee
The non trivial solutions of $\gamma_{4}=\gamma_{2,2}=0$ (as always, up to the choice $\alpha>0$) are 
\be
(\alpha, \beta) = \bigg(\frac{7}{6}, -\frac{2}{3}\bigg), \quad
\bigg(\frac{5}{4}, -\frac{1}{2}\bigg), \quad
\bigg(\frac{11}{6}, -\frac{2}{3}\bigg), \quad
\bigg(\frac{5}{2}, -2\bigg),
\ee
that is precisely the four points $\text{X}_{i}$. The coefficient $\gamma_{2}$ is also an output of the
calculation and its general expression is
\be
\la{6.7}
\gamma_{2} = \frac{(-2 \alpha +\beta -1) (-2 \alpha +\beta +1) (2 \alpha +\beta -1) (2 \alpha
   +\beta +1)}{192 \beta }.
\ee
Finally, comparing (\ref{6.2}) and (\ref{6.3}) we determine the perturbative part of the prepotential
to be, see (\ref{1.3}) 
\be
\wt F\pert_{(\alpha, \beta)}(\nu) = -\beta\,h_{m}\,\log\frac{\nu}{\Lambda}-\beta\log
\prod_{n=1}^{N}\bigg(1-\frac{\nu_{n}^{2}}{\nu^{2}}\bigg),
\ee
that is a nice generalization of (\ref{5.4}). In all cases, we have confirmed that this is 
in full agreement with the explicit evaluation of the general expression (\ref{5.1}). We remark 
once again that it is remarkable that at the special $n$-poles points, it is possible to get such 
a simple expression.

\bigskip

\begin{figure}[!ht]
\begin{center}
\hskip 1cm \includegraphics[scale=0.5]{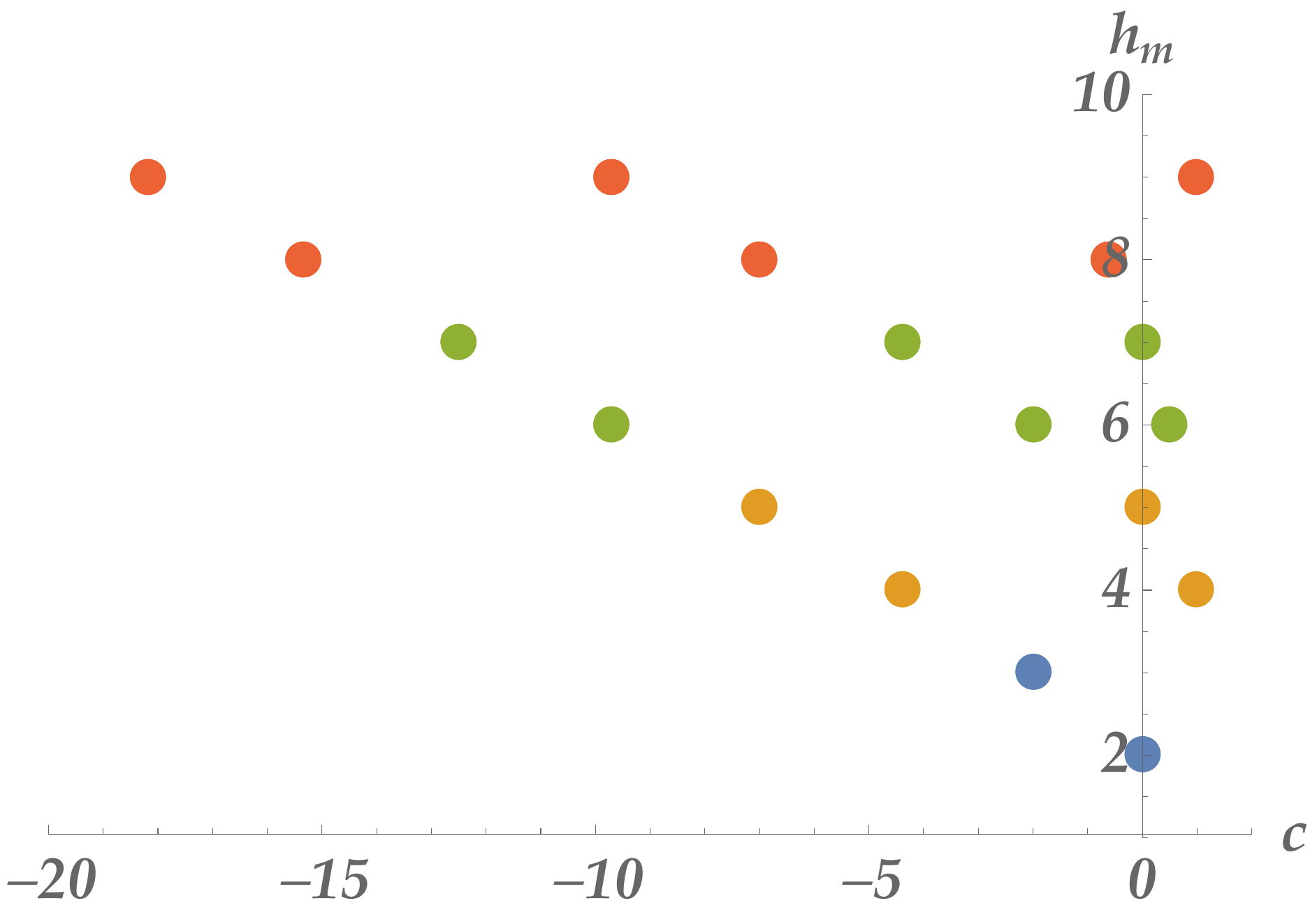}
\caption{Values of  $(c,h_{m})$ for 1, 2, 3, and 4 poles points (blue, orange, green, and 
red).}
\label{fig:chm}
\end{center}
\end{figure}

Performing the above analysis for 2- and 3- poles Nekrasov functions, we find the following
complete results. We have four 1- and seven 2-poles Nekrasov functions that 
are fully characterized by the following tables where we have separated by a horizontal line 
points with differents $(c, h_{m})$
\renewcommand{\arraystretch}{1.5}
\be
\la{6.9}
\begin{array}[t]{|cc|cc|c|c|}
\hline
\alpha & \beta & c & h_{m} & \gamma_{2} & \nu_{1} \\
\hline 
 \frac{5}{2} & -2 & -2 & 3 & -1 & 1 \\
 \frac{5}{4} & -\frac{1}{2} & -2 & 3 & -\frac{1}{4} & \frac{1}{2} \\
 \hline
 \frac{7}{4} & -\frac{3}{2} & 0 & 2 & -\frac{1}{4} & \frac{1}{2} \\
 \frac{7}{6} & -\frac{2}{3} & 0 & 2 & -\frac{1}{9} & \frac{1}{3}\\
 \hline 
\end{array}
\qquad 
\begin{array}[t]{|cc|cc|ccc|cc|}
\hline
\alpha & \beta & c & h_{m} & \gamma_{2} & \gamma_{4} & \gamma_{2,2} & \nu_{1} & \nu_{2}\\
\hline
 4 & -3 & -7 & 5 & -5 & \frac{1}{4} & \frac{15}{4} & 1 & 2 \\
 \frac{4}{3} & -\frac{1}{3} & -7 & 5 & -\frac{5}{9} & \frac{1}{324} & \frac{5}{108} &
   \frac{1}{3} & \frac{2}{3} \\
   \hline
 \frac{13}{4} & -\frac{5}{2} & -\frac{22}{5} & 4 & -\frac{5}{2} & \frac{1}{24} &
   \frac{25}{48} & \frac{1}{2} & \frac{3}{2} \\
 \frac{13}{10} & -\frac{2}{5} & -\frac{22}{5} & 4 & -\frac{2}{5} & \frac{2}{1875} &
   \frac{1}{75} & \frac{1}{5} & \frac{3}{5} \\
   \hline
 2 & -1 & 1 & 4 & -1 & -\frac{1}{12} & \frac{1}{12} & 0 & 1 \\
 \hline
 \frac{11}{4} & -\frac{3}{2} & 0 & 5 & -\frac{5}{2} & -\frac{3}{8} & \frac{15}{16} &
   \frac{1}{2} & \frac{3}{2} \\
 \frac{11}{6} & -\frac{2}{3} & 0 & 5 & -\frac{10}{9} & -\frac{2}{27} & \frac{5}{27} &
   \frac{1}{3} & 1 \\
   \hline
   \end{array}
\ee
Looking for 3-poles points, we identify 12 cases whose full data is collected in the Table in 
(\ref{A.1}). With 4-poles, we found 11 solutions collected in the two
tables (\ref{A.2}) and (\ref{A.3}). Notice that in all presented cases, the parameter  
$\beta$ is always rational negative. This implies that central charge takes the form 
of {\em extended} minimal models
\be
c = 1-6\,\frac{(p-q)^{2}}{p q},
\ee
with coprime integers $p$, $q$. Here, the extension is due to the fact that the minimum value of $p,q$
is one instead of 2, see also Fig.~(\ref{fig:chm}).

\subsection{Constraints from the modular anomaly equation}

An important test of  the expression (\ref{6.2}) is the validity of the 
modular anomaly equation expressing S-duality
 \cite{Billo:2013fi,Billo:2013jba,Billo:2014bja,Billo:2015ria,Billo:2015jta,Billo:2016zbf}.
This is a non-trivial constraint  capturing the dependence on the quasi-modular series $\E_{2}$
and reads \cite{Billo:2013jba}
\be
\la{zxz1}
\frac{\partial\wt F}{\partial \E_{2}}+\frac{1}{12}\,\left(\frac{\partial\wt F}{\partial\nu}\right)^{2}
-\frac{\beta}{12}\,\frac{\partial^{2}\wt F}{\partial\nu^{2}}=0.
\ee
An alternative form is obtained by expanding at large $\nu$ and identifying the coefficients $h_{\ell}$
in 
\be
\la{6.11}
\wt F 
= h_{0}\log\frac{\nu}{\Lambda}-
\sum_{\ell=1}^{\infty}\frac{h_{\ell}}{2^{1-\ell}\,\ell}\frac{1}{\nu^{2\ell}},\qquad h_{0}=-\beta\,h_{m}.
\ee
Then, (\ref{zxz1}) may be written in the equivalent form 
 \be
 \la{6.12}
\frac{\partial h_{\ell}}{\partial \E_{2}} = \frac{\ell}{12}\,\sum_{n=0}^{\ell-1}h_{n}\,h_{\ell-1-n}
+\beta\,\frac{\ell\,(2\,\ell-1)}{12}\,h_{\ell-1}.
\ee
Just to give a simple example, let us consider a 1-pole function and write the $\nu$-dependent part of 
$\wt F$ as 
\be
\la{6.13}
\wt F(\nu) = h_{0}\log\frac{\nu}{\Lambda}-\beta\log
\left(1+\frac{\gamma_{2}\,\E_{2}}{\nu^{2}}\right).
\ee
Imposing (\ref{zxz1}) or (\ref{6.12}), we recover the expression (\ref{6.7}) for $\gamma_{2}$. Besides, we
also get the following constraint between $\alpha$ and $\beta$ (excluding trivial solutions with 
constant prepotential)
\be
\la{6.14}
4\,\alpha^{2}-\beta^{2}+6\,\beta-1=0, \quad\text{or}\quad 4\,\alpha^{2}-\beta^{2}+10\,\beta-1=0.
\ee
The condition (\ref{6.14}) is indeed satisfied by the values in the first table of (\ref{6.9}).
However, there are infinite other pairs $(\alpha, \beta)$ that make (\ref{6.13}) a solution of (\ref{zxz1})
that is not realized in the gauge theory. Of course, this is because (\ref{6.13}) predicts all the higher order
terms in the large $\nu$ expansion and this is correct in comparison with the actual Nekrasov
formulas only for a finite set of values of $\alpha$, and $\beta$. Actually, the 
admissibility of the Ansatz (\ref{6.3}) is definitely non trivial. Anyway, 
we checked the validity of (\ref{zxz1})
for all the solutions we have presented.

As a final comment, we notice that using  the explicit expressions for $\gamma_{2,2}$ and $\gamma_{2}$ it 
turns out that the constraint (\ref{6.14}), up to an  multiplicative factor, can be expressed as the ratio $\gamma_{2,2}/\gamma_{2}$:
\begin{equation}
\frac{\gamma_{2,2}}{\gamma_{2}}  \sim   \left(-4 \alpha ^2+(\beta -10) \beta +1\right) \left(-4 \alpha ^2+(\beta -6) \beta +1\right)  = 0.
\end{equation}
$\gamma_{2,2} = 0$ is of course a necessary condition in the Ansatz (\ref{6.3}) to truncate the sum at $N=1$.
Looking at higher values of $N$, one gets similar constraints for any $N$. In fact, since the modular anomaly equation controls the dependence of $\widetilde{F}$
on $E_2$, the consistency of our Ansatz with (\ref{zxz1})
imposes  relations between all the coefficients of the form $\gamma_{2,X}$ with $\gamma_{X}$. For example, in the 2-poles case we have that
\begin{equation}
\frac{\gamma_{2,2,2}}{\gamma_{2,2}}  \sim   \frac{\gamma_{2,4}}{\gamma_{4}}  \sim   \left(-4 \alpha ^2+(\beta -18) \beta +1\right) \left(-4 \alpha ^2+(\beta -14) \beta +1\right),
\end{equation}
and  the constraint for $\alpha$ and $\beta$ ensuring that $\gamma_{2,2,2} = \gamma_{2,4} = 0$ is
\begin{equation}
-4 \alpha ^2 + \beta^2 -18 \beta +1 = 0  \quad\text{or}\quad -4 \alpha ^2+ \beta^2 -14 \beta +1 = 0.
\end{equation}
Similarly for 3-poles we get
\begin{equation}
\frac{\gamma_{2,2,2,2}}{\gamma_{2,2,2}}  \sim   \frac{\gamma_{2,2,4}}{\gamma_{2,4}}  \sim  \frac{\gamma_{2,6}}{\gamma_{6}}  \sim \left(-4 \alpha ^2+(\beta -26) \beta +1\right) \left(-4 \alpha ^2+(\beta -22) \beta +1\right),
\end{equation}
and so on.

\subsection{A worked out 3-pole example}

To appreciate the result of our analysis, let us consider in some details the first line of (\ref{A.1}).
The instanton partition function for $\text{X}=(\alpha, \beta)=(\frac{11}{2}, -4)$ is up to 12 instantons
{\small
\begin{align}
& \wt Z\inst_{\text{X}}(\nu) = 1-\frac{12 \left(\nu ^2-37\right) q^2}{\nu ^2-9}+\frac{54
   \left(\nu ^4-69 \nu ^2+820\right) q^4}{\left(\nu ^2-9\right)
   \left(\nu ^2-4\right)}\\
   &-\frac{8 \left(11 \nu ^6-1078 \nu
   ^4+28679 \nu ^2-103212\right) q^6}{\left(\nu ^2-9\right)
   \left(\nu ^2-4\right) \left(\nu ^2-1\right)}-\frac{9 \left(11
   \nu ^6-1386 \nu ^4+3339 \nu ^2+239956\right) q^8}{\left(\nu
   ^2-9\right) \left(\nu ^2-4\right) \left(\nu
   ^2-1\right)}\notag \\
   &+\frac{540 \left(\nu ^6-154 \nu ^4+4333 \nu
   ^2-16948\right) q^{10}}{\left(\nu ^2-9\right) \left(\nu
   ^2-4\right) \left(\nu ^2-1\right)}-\frac{2 \left(209 \nu
   ^6-38038 \nu ^4+1558361 \nu ^2-15934932\right)
   q^{12}}{\left(\nu ^2-9\right) \left(\nu ^2-4\right) \left(\nu
   ^2-1\right)}\notag \\
   &-\frac{648 \left(\nu ^6-210 \nu ^4+11109 \nu
   ^2-61860\right) q^{14}}{\left(\nu ^2-9\right) \left(\nu
   ^2-4\right) \left(\nu ^2-1\right)}+\frac{54 \left(11 \nu
   ^6-2618 \nu ^4+225659 \nu ^2-2870732\right) q^{16}}{\left(\nu
   ^2-9\right) \left(\nu ^2-4\right) \left(\nu
   ^2-1\right)}\notag \\
   &+\frac{4 \left(209 \nu ^6-55594 \nu ^4+3095981 \nu
   ^2-36365076\right) q^{18}}{\left(\nu ^2-9\right) \left(\nu
   ^2-4\right) \left(\nu ^2-1\right)}+\frac{96 \left(11 \nu
   ^6-3234 \nu ^4+58779 \nu ^2+2749204\right) q^{20}}{\left(\nu
   ^2-9\right) \left(\nu ^2-4\right) \left(\nu
   ^2-1\right)}\notag \\
   & -\frac{216 \left(19 \nu ^6-6118 \nu ^4+358351 \nu
   ^2-3592972\right) q^{22}}{\left(\nu ^2-9\right) \left(\nu
   ^2-4\right) \left(\nu ^2-1\right)}
   -\frac{\left(209 \nu
   ^6-73150 \nu ^4+18811121 \nu ^2-150221700\right)
   q^{24}}{\left(\nu ^2-9\right) \left(\nu ^2-4\right) \left(\nu
   ^2-1\right)}+\dots.
   \end{align}
   }
This is far more involved than (\ref{3.1}) and a brute force guess would not be possible. 
Nevertheless, it is a straightforward calculation to check that this is the expansion of 
\be
\wt Z\inst_{\text{X}}(\nu) = [q^{-\frac{1}{12}}\eta(\tau)]^{12}\,
\frac{\nu^{6}-14\, \E_2 \nu ^4+\left(\frac{140}{3}\, \E_2^2+\frac{7}{3}\,
\E_4\right) \nu ^2-\frac{280}{9}\, \E_2^3-\frac{14}{3} \,\E_2\,
   \E_4-\frac{2}{9}\, \E_6}{\left(\nu ^2-9\right)
   \left(\nu ^2-4\right) \left(\nu ^2-1\right)},
\ee
in agreement with the data in (\ref{A.1}).
The perturbative part of the prepotential is computed from (\ref{5.1}) by expanding at large $a$. 
This gives, in the $\nu$ variable
\be
\widetilde F\pert_{\rm X}(\nu) = 28\,\log\frac{\nu}{\Lambda}-\frac{56}{\nu
   ^2}-\frac{196}{\nu ^4}-\frac{3176}{3 \nu ^6}-\frac{6818}{\nu
   ^8}-\frac{240296}{5 \nu ^{10}}-\frac{1071076}{3 \nu
   ^{12}}-\frac{2742488}{\nu ^{14}}+\dots,
\ee
and this is indeed the large $\nu$ expansion of 
\be
\widetilde F\pert_{\rm X}(\nu) = 28\,\log\frac{\nu}{\Lambda}+4\,\log\bigg[
\bigg(1-\frac{1}{\nu^{2}}\bigg)\,\bigg(1-\frac{4}{\nu^{2}}\bigg)\,\bigg(1-\frac{9}{\nu^{2}}\bigg)
\bigg],
\ee
where $28 = -\beta\,h_{m} = -(-4)\times 7$.  Hence, the full quantum prepotential is in this case
\be
\wt F = 28\,\log\frac{\nu}{\Lambda}+4\,\log\bigg[
1-\frac{14}{\nu^{2}}\,\E_{2}+\frac{7}{3\,\nu^{4}}\,(20\,\E_{2}^{2}+\E_{4})-\frac{2}{9\,\nu^{6}}\,
(140\,\E_{2}^{3}+21\,\E_{2}\,\E_{4}+\E_{6})
\bigg],
\ee
and one can check that (\ref{zxz1}) is satisfied.
%Equivalently, the large $\nu$ expansion of the total 
%prepotential is  (dropping $\nu$-independent terms)
%\be
%\begin{split}
%\wt F_{\rm X}(\nu) &= 28\,\log\frac{\nu}{\Lambda}+4\,\log\bigg[
%1+\frac{-14\, \E_2 \nu ^4+\left(\frac{140}{3}\, \E_2^2+\frac{7}{3}\,
%\E_4\right) \nu ^2-\frac{280}{9}\, \E_2^3-\frac{14}{3} \,\E_2\,
%   \E_4-\frac{2}{9}\, \E_6}{\nu^{6}}\bigg] \\
%   &= 28\,\log\frac{\nu}{\Lambda} -\frac{56\, \E_2}{\nu ^2}+\frac{28\, \left(\E_4-22\, \E_2^2\right)}{3 \nu
%   ^4}-\frac{8 \left(1316\, \E_2^3-126\, \E_4 \E_2+\E_6\right)}{9 \nu
%   ^6}\\
%   &-\frac{14 \left(5096\, \E_2^4-728\, \E_4 \E_2^2+8\, \E_6 \E_2+7\,
%   \E_4^2\right)}{9 \nu ^8}\\
%   &-\frac{56 \left(141232\, \E_2^5-26320\, \E_4
%   \E_2^3+320\, \E_6 \E_2^2+630\, \E_4^2 \E_2-5 \E_4 \E_6\right)}{135 \nu
%   ^{10}}+\dots.
%\end{split}
%\ee
%Reading the coefficients $h_{\ell}$ in the modular anomaly equation, we get
%\be
%\begin{split}
%h_{0} &= 28, \quad
%h_{1} = 56 \E_2, \quad
%h_{2} = \frac{28}{3} \left(22\, \E_2^2-\E_4\right), \quad
%h_{3} = \frac{2}{3}
%   \left(1316\, \E_2^3-126\, \E_4 \E_2+\E_6\right),\\
%h_{4} &= \frac{7}{9} \left(5096
%   \,\E_2^4-728 \,\E_4 \E_2^2+8 \,\E_6 \E_2+7 \,\E_4^2\right),\\
%h_{5} &= \frac{7}{54}
%   \left(141232 \,\E_2^5-26320 \,\E_4 \E_2^3+320 \,\E_6 \E_2^2+630 \,\E_4^2
%   \E_2-5 \,\E_4 \E_6\right),
%\end{split}
%\ee
%and so on. One can check that (\ref{6.12}) is satisfied.

\section{Predictions for the torus 1-block}
\la{sec:block}

We have already seen that AGT implies  explicit expressions for 
 special torus blocks that we write stripping off the large $h$ dominant term
 \be
\mc F^{h}_{h_{m}}(q, c) = \frac{q^{\frac{1}{12}}}{\eta(\tau)}\,\mc H^{h}_{h_{m}}(q,c).
\ee
From the 1-pole partition functions, we have obtained
\be
\begin{split}
\mc H^{h}_{2}(q, 0) =1+\frac{1-\E_2}{24 h},\quad
\mc H^{h}_{3}(q, -2) = 1+\frac{1-\E_2}{8 h}.
\end{split}
\ee
From the 2-poles partition functions, we get similar expressions
\be
\begin{split}
\mc H^{h}_{5}(q,-7) &= 1+\frac{15 \E_2^2+\E_4-80 \E_2 (3 \,h+1)+16 (15 \,h+4)}{144 \,h (4 \,h+1)}, \\
\mc H^{h}_{4}(q,-\tfrac{22}{5}) &= 1+\frac{25 \E_2^2+2 \E_4-30 \E_2 (40 \,h+9)+3 (400 \,h+81)}{960 \,h (5 \,h+1)}, \\
\mc H^{h}_{4}(q,1) &= 1+\frac{\E_2^2-\E_4-48 \E_2 \,h+48 \,h}{48 \,h (4 \,h-1)}, \\
\mc H^{h}_{5}(q,0) &= 1+\frac{3 \left(5 \E_2^2-2 \E_4\right)-10 \E_2 (24 \,h+1)+240 \,h+1}{192 \,h (3 \,h-1)}.
\end{split}
\ee
Notice that there are multiple entries in the tables (\ref{6.9}) and (\ref{A.1})
with the same value of $(c,h_{m})$, but different $(\alpha, \beta)$. 
Nevertheless, the associated  block is consistently the same as soon as 
$\nu$ is expressed in terms of $h$.
From the 3-poles partition functions, we get
\begin{align}
\mc H^{h}_{7}(q,-\tfrac{25}{2}) &= 1+\frac{1}{1152 \,h (2 \,h+1) (16 \,h+5)}\,\bigg[
-2 (140 \,\E_2^3+21 \,\E_4 \,\E_2+\,\E_6)+9 (3584 \,h^2+3248
   \,h+729)\notag\\
   &-126 \,\E_2 (16 \,h+9)^2+21 (20 \,\E_2^2+\,\E_4) (16 \,h+9)
\bigg], \notag \\
\mc H^{h}_{6}(q,-\tfrac{68}{7}) &= 1+\frac{1}{32256 \,h (7 \,h+2) (7 \,h+3)}\,\bigg[
-1715 \,\E_2^3-294 \,\E_4 \,\E_2-16 \,\E_6\notag \\
&+9 (109760 \,h^2+83496 \,h+15625)-315
   \,\E_2 (56 \,h+25)^2+63 (35 \,\E_2^2+2 \,\E_4) (56 \,h+25)
\bigg], \notag \\
\mc H^{h}_{7}(q,-\tfrac{22}{5}) &= 1+\frac{1}{23040 \,h (5 \,h-2) (5 \,h+1)}\,\bigg[
-5 (875 \,\E_2^3-294 \,\E_4 \,\E_2-176 \,\E_6)\notag \\
&+9 (56000 \,h^2+14840
   \,h+729)-315 \,\E_2 (40 \,h+9)^2+21 (125 \,\E_2^2-14 \,\E_4) (40
   \,h+9)
\bigg], \notag \\
\mc H^{h}_{6}(q,-2) &= 1+\frac{1}{72 \,h (8 \,h-3) (8 \,h+1)}\,\bigg[
-5 \,\E_2^3+3 \,\E_4 \,\E_2+2 \,\E_6-45 \,\E_2 (8 \,h+1)^2\notag \\
&+9 (5 \,\E_2^2-\,\E_4) (8
   \,h+1)+9 (8 \,h+1) (40 \,h+1)
\bigg], \notag \\
\mc H^{h}_{6}(q,\tfrac{1}{2}) &= 1+\frac{1}{3456 \,h (2 \,h-1) (16 \,h-1)}\,\bigg[
-2 (60 \,\E_2^3-51 \,\E_4 \,\E_2+41 \,\E_6)+69120 \,h^2\notag \\
&-30 \,\E_2 (48 \,h+1)^2+3
   (60 \,\E_2^2-17 \,\E_4) (48 \,h+1)-3312 \,h+1
\bigg], \notag \\
\mc H^{h}_{7}(q,0) &= 1+\frac{1}{216 \,h (8 \,h-5) (8 \,h-1)}\,\bigg[
-105 \,\E_2^3+63 \,\E_4 \,\E_2-22 \,\E_6+12096 \,h^2\notag \\
&-21 \,\E_2 (24 \,h+1)^2+21 (5
   \,\E_2^2-\,\E_4) (24 \,h+1)-1008 \,h+1
\bigg].
\end{align}
All these expressions can be re-expanded at small $q$ and compared with the recursion relations for the 
toroidal block with full agreement. It would be  interesting to understand 
why such closed expressions are obtained at the special $(c, h_{m})$ points, for instance, 
by extending the results of  
\cite{Gaberdiel:2008ma,Gaberdiel:2009vs}.

\section*{Acknowledgments}
We  thank A. Zein Assi for important comments and D. Fioravanti for clarifying discussions.

\newpage
\appendix
\section{3- and 4-poles Nekrasov functions data}

The 3-poles Nekrasov functions are 
\be
\la{A.1}
\begin{array}{|cc|cc|cccccc|ccc|}
\hline
\alpha & \beta & c & h_{m} & \gamma_{2} & \gamma_{4} & \gamma_{2,2} & 
\gamma_{6} & \gamma_{2,4} & \gamma_{2,2,2} & \nu_{1} & \nu_{2} & \nu_{3} \\
\hline
\frac{11}{2} & -4 & -\frac{25}{2} & 7 & -14 & \frac{7}{3} & \frac{140}{3} &
   -\frac{2}{9} & -\frac{14}{3} & -\frac{280}{9} & 1 & 2 & 3 \\
 \frac{11}{8} & -\frac{1}{4} & -\frac{25}{2} & 7 & -\frac{7}{8} & \frac{7}{768} &
   \frac{35}{192} & -\frac{1}{18432} & -\frac{7}{6144} & -\frac{35}{4608} &
   \frac{1}{4} & \frac{1}{2} & \frac{3}{4} \\
   \hline
 \frac{19}{4} & -\frac{7}{2} & -\frac{68}{7} & 6 & -\frac{35}{4} & \frac{7}{8} &
   \frac{245}{16} & -\frac{1}{36} & -\frac{49}{96} & -\frac{1715}{576} & \frac{1}{2} &
   \frac{3}{2} & \frac{5}{2} \\
 \frac{19}{14} & -\frac{2}{7} & -\frac{68}{7} & 6 & -\frac{5}{7} & \frac{2}{343} &
   \frac{5}{49} & -\frac{16}{1058841} & -\frac{2}{7203} & -\frac{5}{3087} &
   \frac{1}{7} & \frac{3}{7} & \frac{5}{7} \\
   \hline
 \frac{17}{4} & -\frac{5}{2} & -\frac{22}{5} & 7 & -\frac{35}{4} & -\frac{49}{24} &
   \frac{875}{48} & \frac{55}{36} & \frac{245}{96} & -\frac{4375}{576} & \frac{1}{2} &
   \frac{3}{2} & \frac{5}{2} \\
 \frac{17}{10} & -\frac{2}{5} & -\frac{22}{5} & 7 & -\frac{7}{5} & -\frac{98}{1875} &
   \frac{7}{15} & \frac{176}{28125} & \frac{98}{9375} & -\frac{7}{225} & \frac{1}{5} &
   \frac{3}{5} & 1 \\
   \hline
 \frac{7}{2} & -2 & -2 & 6 & -5 & -1 & 5 & \frac{2}{9} & \frac{1}{3} & -\frac{5}{9} &
   0 & 1 & 2 \\
 \frac{7}{4} & -\frac{1}{2} & -2 & 6 & -\frac{5}{4} & -\frac{1}{16} & \frac{5}{16} &
   \frac{1}{288} & \frac{1}{192} & -\frac{5}{576} & 0 & \frac{1}{2} & 1 \\
   \hline
 \frac{17}{6} & -\frac{4}{3} & \frac{1}{2} & 6 & -\frac{10}{3} & -\frac{17}{27} &
   \frac{20}{9} & -\frac{82}{729} & \frac{34}{243} & -\frac{40}{243} & \frac{1}{3} &
   \frac{2}{3} & \frac{5}{3} \\
 \frac{17}{8} & -\frac{3}{4} & \frac{1}{2} & 6 & -\frac{15}{8} & -\frac{51}{256} &
   \frac{45}{64} & -\frac{41}{2048} & \frac{51}{2048} & -\frac{15}{512} & \frac{1}{4}
   & \frac{1}{2} & \frac{5}{4} \\
   \hline
 \frac{13}{4} & -\frac{3}{2} & 0 & 7 & -\frac{21}{4} & -\frac{21}{16} & \frac{105}{16}
   & -\frac{11}{32} & \frac{63}{64} & -\frac{105}{64} & \frac{1}{2} & 1 & 2 \\
 \frac{13}{6} & -\frac{2}{3} & 0 & 7 & -\frac{7}{3} & -\frac{7}{27} & \frac{35}{27} &
   -\frac{22}{729} & \frac{7}{81} & -\frac{35}{243} & \frac{1}{3} & \frac{2}{3} &
   \frac{4}{3} \\
   \hline
\end{array}
\ee
The 4-poles data is split in the following two tables. The first contains the values of
$(c, h_{m})$ and the poles positions for each pair $(\alpha, \beta)$.
\begin{eqnarray}
\la{A.2}
\footnotesize
\begin{array}{|cc|cc|cccc|}
\hline 
 \alpha  & \beta  & c & h_m & \nu _1 & \nu _2 & \nu _3 & \nu _4 \\
 \hline 
 7 			& -5 			& -\frac{91}{5} 		& 9 	& 1 			& 2 			& 3 			& 4 \\
 \frac{7}{5} 	& -\frac{1}{5} 	& -\frac{91}{5} 		& 9 	& \frac{1}{5} 	& \frac{2}{5} 	& \frac{3}{5} 	& \frac{4}{5} \\
 \hline 
 \frac{25}{4} 	& -\frac{9}{2} 	& -\frac{46}{3} 		& 8 	& \frac{1}{2} 	& \frac{3}{2} 	& \frac{5}{2} 	& \frac{7}{2} \\
 \frac{25}{18} 	& -\frac{2}{9} 	& -\frac{46}{3} 		& 8 	& \frac{1}{9} 	& \frac{1}{3} 	& \frac{5}{9} 	& \frac{7}{9} \\
 \hline  
 \frac{23}{4} 	& -\frac{7}{2} 	& -\frac{68}{7} 		& 9 	& \frac{1}{2} 	& \frac{3}{2} 	& \frac{5}{2} 	& \frac{7}{2} \\
 \frac{23}{14} 	& -\frac{2}{7} 	& -\frac{68}{7} 		& 9 	& \frac{1}{7} 	& \frac{3}{7} 	& \frac{5}{7} 	& 1 \\
 \hline 
 5 			& -3 			& -7 				& 8 	& 0 			& 1 			& 2 			& 3 \\
 \frac{5}{3} 	& -\frac{1}{3} 	& -7 				& 8 	& 0 			& \frac{1}{3} 	& \frac{2}{3} 	& 1 \\
 \hline 
 \frac{11}{3} 	& -\frac{5}{3} 	& -\frac{3}{5} 		& 8 	& \frac{1}{3} 	& \frac{2}{3} 	& \frac{4}{3} 	& \frac{7}{3} \\
 \frac{11}{5} 	& -\frac{3}{5} 	& -\frac{3}{5} 		& 8 	& \frac{1}{5} 	& \frac{2}{5}	& \frac{4}{5} 	& \frac{7}{5} \\
 \hline 
 3 			& -1 			& 1 				& 9 	& 0 			& 1 			& 1			& 2 \\
 \hline 
\end{array}
\end{eqnarray}
The next table lists the values of the $\gamma$-coefficients.
\begin{eqnarray}
\la{A.3}
\footnotesize
\begin{array}{|cc|c|cc|ccc|cccc|}
\hline 
 \alpha  & \beta  & \gamma _2 & \gamma _4 & \gamma _{2,2} & \gamma _6 & \gamma _{2,4} & \gamma _{2,2,2} & \gamma _{4,4} & \gamma _{2,6} & \gamma _{2,2,4} & \gamma _{2,2,2,2} \\
 \hline 
 7 			& -5 				& -30 		& \frac{21}{2} & \frac{525}{2} & -\frac{10}{3} & -\frac{175}{2} & -\frac{4375}{6} & \frac{41}{16} & \frac{25}{3} & \frac{875}{8} & \frac{21875}{48} \\
 \frac{7}{5} 	& -\frac{1}{5} 		& -\frac{6}{5} 	& \frac{21}{1250} & \frac{21}{50} & -\frac{2}{9375} & -\frac{7}{1250} & -\frac{7}{150} & \frac{41}{6250000} & \frac{1}{46875} & \frac{7}{25000} & \frac{7}{6000} \\
 \hline 
 \frac{25}{4}	 & -\frac{9}{2} 		& -21 		& \frac{21}{4} & \frac{945}{8} & -1 & -\frac{189}{8} & -\frac{2835}{16} & \frac{15}{64} & \frac{3}{4} & \frac{567}{64} & \frac{8505}{256} \\
 \frac{25}{18} 	& -\frac{2}{9} 		& -\frac{28}{27}	& \frac{28}{2187} & \frac{70}{243} & -\frac{64}{531441} & -\frac{56}{19683} & -\frac{140}{6561} & \frac{20}{14348907} & \frac{64}{14348907} & \frac{28}{531441} & \frac{35}{177147} \\
 \hline 
 \frac{23}{4} 	& -\frac{7}{2} 		& -21 		& -\frac{21}{4} & \frac{1029}{8} & \frac{53}{3} & \frac{245}{8} & -\frac{12005}{48} & -\frac{553}{64} & -\frac{371}{12} & -\frac{1715}{64} & \frac{84035}{768} \\
 \frac{23}{14} 	& -\frac{2}{7} 		& -\frac{12}{7} 	& -\frac{12}{343} & \frac{6}{7} & \frac{3392}{352947} & \frac{40}{2401} & -\frac{20}{147} & -\frac{316}{823543} & -\frac{3392}{2470629} & -\frac{20}{16807} & \frac{5}{1029} \\
 \hline 
 5 			& -3 				& -14 		& -\frac{7}{2} & \frac{105}{2} & 6 & \frac{21}{2} & -\frac{105}{2} & -\frac{15}{16} & -3 & -\frac{21}{8} & \frac{105}{16} \\
 \frac{5}{3} 	& -\frac{1}{3} 		& -\frac{14}{9} 	& -\frac{7}{162} & \frac{35}{54} & \frac{2}{243} & \frac{7}{486} & -\frac{35}{486} & -\frac{5}{34992} & -\frac{1}{2187} & -\frac{7}{17496} & \frac{35}{34992} \\
 \hline 
 \frac{11}{3} 	& -\frac{5}{3} 		& -\frac{70}{9} 	& -\frac{133}{54} & \frac{875}{54} & -\frac{610}{729} & \frac{665}{162} & -\frac{4375}{486} & \frac{6667}{34992} & \frac{1525}{6561} & -\frac{3325}{5832} & \frac{21875}{34992} \\
 \frac{11}{5} 	& -\frac{3}{5} 		& -\frac{14}{5} 	& -\frac{399}{1250} & \frac{21}{10} & -\frac{122}{3125} & \frac{1197}{6250} & -\frac{21}{50} & \frac{20001}{6250000} & \frac{61}{15625} & -\frac{1197}{125000} & \frac{21}{2000} \\
 \hline 
 3 			& -1 				& -6 			& -\frac{3}{2} & \frac{21}{2} & -\frac{2}{3} & \frac{5}{2} & -\frac{35}{6} & -\frac{7}{16} & \frac{1}{3} & -\frac{5}{8} & \frac{35}{48} \\
 \hline 
\end{array}
\end{eqnarray}

\section{On the structure of higher instanton  Nekrasov functions}

From the analysis in the main text, it is clear that in all the cases  the $k$-instanton 
contribution to $\widetilde{F}^{\text{inst}}(\nu)$ is a rational function of $\nu$: 
in particular in the 1-pole cases the structure of the $\widetilde{F}_{X_i, k}^{\text{inst}}(\nu)$ is in close analogy with the 
Nekrasov-Shatashvili limit studied in \cite{Beccaria:2016wop}. It can be instructive to repeat some of the 
steps of the analysis of \cite{Beccaria:2016wop} and highlight various differences.
Taking the 1-pole point
$X_1$ case as example,  we have 
\begin{equation}
\widetilde{F}_{X_1, k}^{\text{inst}}(\nu) = \frac{P_{k}(\nu)}{\left(\nu^2- 1 \right)^k},
\end{equation}
where  $P_{k}(\nu)$ are even polynomials of degree $2 k$. 
The first cases are explicitly
\begin{align}
P_{1}(\nu) =&  -8 \left(\nu ^2-7\right), \\
P_{2}(\nu) =&  -12 \left(\nu ^4-14 \nu ^2+61\right), \notag \\
P_{3}(\nu) =&  -\frac{32}{3} \left(\nu ^6-21 \nu ^4+363 \nu ^2-1207\right), \notag \\
P_{4}(\nu) =&  -2 \left(7 \nu ^8-196 \nu ^6+5442 \nu ^4-51796 \nu ^2+129487\right), \notag \\
P_{5}(\nu) =&  -\frac{48}{5} \left(\nu ^{10}-35 \nu ^8+2410 \nu ^6-44470 \nu ^4+289205 \nu ^2-578887\right), \notag \\
P_{6}(\nu) =&  -16 \left(\nu ^{12}-42 \nu ^{10}+2715 \nu ^8-79580 \nu ^6+948615 \nu ^4-4655130 \nu ^2+7764733\right), \notag\\
 & \dots\notag
\end{align}
In the cases treated in \cite{Beccaria:2016wop}, the expansion around the poles revealed a 
\emph{selection rule} forbidding even poles, 
{\em i.e.} the functions $\widetilde{F}_{\text{NS},k}^{\text{inst}}(\nu)$ have always the form
\begin{equation}
\widetilde{F}_{\text{NS}, k}^{\text{inst}}(\nu)  = \frac{d_1^{(k)}}{(\nu -1)^{2 k -1}} + \frac{d_{2}^{(k)}}{(\nu-1)^{2k-3}} + \dots + \frac{d_{k}^{(k)}}{(\nu -1)} + \text{regular}.
\end{equation}
While the structure is extremely similar,  the \emph{selection rule} is no longer true here. 
At the point $X_1$, the Laurent  
expansion of $\widetilde{F}_{X_1,k}^{\text{inst}}(\nu)$ 
around $\nu=1$  has the generic form 
\begin{equation}
\widetilde{F}_{X1, k}^{\text{inst}}(\nu) = \frac{d_0^{(k)}}{(\nu -1)^k} + \frac{d_1^{(k)}}{(\nu -1)^{k-1}} + \dots + \frac{d_{k-1}^{(k)}}{(\nu -1)} + \text{regular},
\end{equation}
where all the coefficients are non vanishing. The same is true for all the other three 1-pole cases  $X_2, X_3, X_4$.
As in \cite{Beccaria:2016wop}, 
we can rewrite the $k$-instanton functions in terms of the $d_p^{(k)}$ in the exact form 
\begin{equation}
\widetilde{F}_{X_1,k}^{\text{inst}}(\nu)  = c_k +  \sum_{p=0}^{k-1}{ d_p^{(k)} \left( \frac{(-1)^{k-p}}{(\nu - 1)^{k-p}} + \frac{1}{(\nu +1)^{k-p}}   \right)}.
\end{equation}
The coefficients $c_k$  capture the $\nu$ independent part ({\em i.e.} the term proportional to the
logarithms of the Dedekind function. The coefficients $d_p^{(k)}$  
can be  obtained from the expansion of the exact expressions of $\widetilde{F}_{X_i,k}^{\text{inst}}(\nu)$. For $X_1$ they read
\begin{align}
d_0^{(k)} = &\,\, 2 \frac{12^k}{k},\notag \\
d_1^{(k)} = &\,\, d_0^{(k)}\, \times\frac{3 k}{4},\notag \\
d_2^{(k)} = &\,\, d_0^{(k)}\, \times \frac{1}{288} k (81 k-19), \\
d_3^{(k)} = &\,\, d_0^{(k)}\, \times \frac{k \left(243 k^2-171 k+470\right)}{3456},\notag \\
d_4^{(k)} = &\,\, d_0^{(k)}\, \times \frac{k \left(2187 k^3-3078 k^2+17281 k-9662\right)}{165888},\notag \\
& \dots\notag
\end{align}
and so on.

\bibliography{N2-Biblio}
\bibliographystyle{JHEP}

\end{document}